# Bayesian Optimization of Metastable Nickel Formation During the Spontaneous Crystallization under Extreme Conditions

Sina Malakpour Estalaki[1], Tengfei Luo[1,2,*], Khachatur V. Manukyan[3]

[1]Department of Aerospace and Mechanical Engineering, University of Notre Dame, Notre Dame, Indiana, 46556, USA

[2]Department of Chemical and Biomolecular Engineering, University of Notre Dame, Notre Dame, Indiana, 46556, USA

[3]Nuclear Science Laboratory, Department of Physics, University of Notre Dame, Notre Dame, Indiana 46556, United States

[*]Corresponding author: tluo@nd.edu

# Abstract

Spontaneous crystallization of metals under extreme conditions is a unique phenomenon occurring under far-from-equilibrium conditions that could enable the development of revolutionary and disruptive metastable metals with unusual properties. In this work, the formation of the hexagonal close-packed Nickel (hcp-Ni) metastable phase during spontaneous crystallization is studied using non-equilibrium molecular dynamics (MD) simulations, with the goal of maximizing the fraction of this metastable phase in the final state. We employ Bayesian Optimization (BO) with the Gaussian Processes (GP) regression as the surrogate model to maximize the hcp-Ni phase fraction, where temperature and pressure are control variables. MD simulations provide data for training the GP model, which is then used with BO to predict the next simulation condition. Such a BO-guided active learning leads to a maximum hcp-Ni fraction of 43.38% in the final crystalized phase within 40 iterations when a face-centered cubic (fcc) crystallite serves as the seed for crystallization from the amorphous phase. When an hcp seed is used, the maximum hcp-Ni fraction in the final crystal increases to 58.25% with 13 iterations. This study shows the promise of using BO to identify the process conditions that can maximize the rare phases. This method can also be generally applicable to process optimization to achieve target material properties.

# 1. Introduction

Metastable crystalline materials often exhibit superior properties to their stable counterparts. The most well-known example is diamond, which possesses superior mechanical properties compared to the energetically more stable graphite [1,2]. Many transition metals such as nickel (Ni), cobalt (Co) and tungsten (W) can also exist in metastable forms [3–10]. For example, W occurs in two crystalline phases: α and β [7,11,12]. The former has a body-centered cubic (bcc) structure and is thermodynamically more stable in bulk form. The β phase (also called A15) is metastable with a cubic crystal structure, containing eight atoms per unit cell. The α and β phases can coexist under ambient conditions in thin films. These two phases exhibit different properties [7,13,14]. α-W has a higher electrical conductivity [15,16], while β-W was reported to exhibit superior mechanical properties and a giant spin Hall effect due to spin-orbit torques [14,17]. Thin films containing metastable bcc-Co and hexagonal closed packed (hcp) Ni phases also show unusual magnetic and electric properties [4,9,18,19]. For example, hcp-Ni exhibits unusual magnetic and mechanical properties, as well as high catalytic activity [6–10,15,20].

Our previous experimental and theoretical works [21,22] show that a novel spontaneous crystallization method [21,22] can enable the production of metastable hcp-Ni from amorphous Ni (a-Ni). Molecular Dynamics (MD) simulations show that a crystalline Ni seed can initiate a crystallization process in a cluster of a-Ni nanoparticles at 800 K [21]. The self-sustaining crystallization wave traveling through the cluster leads to metastable hcp phase at the early stage of crystallization, but they transition into the more stable face-centered cubic (fcc) phase eventually. These simulation results were confirmed by experimental investigation: the localized heating of the compressed a-Ni nanoparticles (produced by liquid-phase reduction of nickel nitrate) triggers a self-sustaining crystallization wave that propagates along the sample, eventually producing the stable fcc phase [21].

Some studies have been done on the crystallization of Ni using MD simulations. Nguyen *et al.* utilized MD simulations to study the influence of the heating and cooling rates, number of particles, temperature and relaxation time on the microscopic structure, phase transitions and dynamics of crystallization in four model systems containing different number of Ni atoms. Based on the common neighbor analysis (CNA), they discovered the coexistence of amorphous and crystalline phases during the crystallization process [23]. Louzguine-Luzgin *et al.* studied the vitrification and crystallization processes in liquid Ni using MD simulations. In their work, the glass transition was monitored using the radial distribution functions and cluster analysis. They analyzed the crystallization kinetics under isothermal conditions by monitoring density and energy variation as a function of time and found that the temperature corresponding to the minimum incubation period in the time-temperature-transformation diagram is 700–900 K [24]. Recently, we investigated the mechanism of spontaneous crystallization process in a-Ni nanoparticles to produce hcp-Ni [22]. Detailed MD simulations predict that hcp-Ni can be stabilized by tuning the amount and positions of crystalline seeds incorporated into the a-Ni nanoparticles. However, these simulations showed that the most critical parameter is temperature, with the optimal range of 700-900 K. The fraction of hcp-Ni in the resulting crystalline product is ~ 20%. Rapid heating and cooling experiments of a-Ni nanoparticles prove the MD simulation results and enable the stabilization of a sizable quantity of hcp-Ni phase. The resulting hcp/fcc-Ni materials are shown to have superior mechanical properties compared to pure fcc-Ni.

Other than Ni, researchers also studied crystallization and self-propagating waves in other metals. Chui *et al.* used a topological method based on planar graphs to analyze the crystalline structure of gold nanoclusters formed by quenching from the melt by MD [25]. Mahata *et al.* investigated the homogeneous nucleation from Al melt by large scale MD simulations. They found that the main crystalline phase was identified as fcc, but a hcp and an amorphous solid phases were also detected [26]. As can be seen from these previous studies, MD simulation is a valuable tool to facilitate the understanding of the metastable phase from a molecular level. However, each of such simulations can be time-consuming. Therefore, optimizing the amount of the metastable phase by exhaustively trying different conditions (*e.g.*, temperature and pressure) can be insurmountable using MD simulations and there can be no guarantee of success.

Recently, the advent of black-box optimizations like Bayesian Optimization (BO) and Machine Learning (ML) models led to the novel material design and discovery with superior properties [27–34]. Terayama *et al.* gave an overview of recent studies regarding automated discovery, design, and optimization based on black-box optimization methods including BO [35]. Diwale *et al.* utilized an augmented BO to promote the nucleation of polyethylene crystals with noisy measurements from non-equilibrium MD simulations [36]. Solomou *et al.* used a multi-objective BO approach to efficiently discover the precipitation-strengthened NiTi shape memory alloys with up to three desired properties [37].

In this work, we use MD simulations to model the spontaneous crystallization process of a-Ni under different temperatures and pressures. The final fraction of the metastable hcp phase from each simulation is recorded, which provides data for training a Gaussian Process (GP) machine learning model. We then utilize BO with the GP to suggest the next simulation conditions (temperature and pressure) based on the expected improvement acquisition function, which balances exploration and exploitation in the parameter space. A new MD simulation is then performed for the BO-suggested conditions and the calculated hcp-Ni fraction is added to the dataset to re-train the GP model to start a new iteration. Such a BO-guided active learning leads to a maximum hcp-Ni fraction of 43.38% in the final crystalized phase within 40 iterations when a fcc crystallite serves as the seed for crystallization from the amorphous phase. When a hcp seed is used, the maximum hcp-Ni fraction in the final crystal increases to 58.25% with 13 iterations. This study shows the promise of using BO to identify the process conditions that can maximize the rare phases. This method can also be generally applicable to process optimization to achieve target material properties.

## 2. Methodology

*2.1. Crystallization simulations*

In order to study the crystallization of a-Ni, it is needed to first generate a-Ni structure. We did this by melting and quenching a crystal Ni structure, which can be either fcc or hcp. For fcc and hcp crystals, the MD simulation box sizes are chosen to be 10×10×45 and 15×8.66×45 unit cells in size containing 18,000 and 23,625 Ni atoms, respectively. Each of these single crystals is thermalized at 300 K under 1 bar for 100 ps. Then it is heated up from 300 K to 3000 K under 1 bar in the NPT ensemble for 1000 ps to be melted, and the molten structure is equilibrated at 3000 K for 100 ps. The structure is then quenched to 300 K within 100 ps to form the a-Ni structure. The structure is further equilibrated at 300 K and different pressures for 100 ps to get prepared for the crystallization simulation. After that, a crystalline seed (hcp or fcc) is placed in contact with the a-Ni at one end of the simulation domain, and after building the combined structure, an energy minimization is performed on it. Finally, the combined structure is simulated at the target $T_i$ and $P_i$ for 1250 ps. **Figures 1a** and **1b** show the amorphization process of fcc-Ni and the crystallization process of a-Ni, respectively. For all cases, the isothermal and isobaric MD simulations are performed using Nosé−Hoover thermostat and barostat with relaxation time scales of 0.1 and 1 ps, respectively. **Figure S1** in **section S1** of the Supporting Information (SI) shows the evolution of pressure and temperature with time for the pairs of input process conditions equal to (800 K, 10 bar) and (1400 K, 200000 bar) corresponding to the a-Ni with fcc and hcp seeds, respectively. This demonstrates that pressure and temperature values are well-controlled in the system during the crystallization of a-Ni under NPT ensemble.

All simulations performed are using the Large-scale Atomic/Molecular Massively Parallel Simulator (LAMMPS) Package [38,39]. The embedded-atom method potential is used [21,22,40,41]. The fraction of the hcp-Ni phase is calculated based on the Polyhedral Template Matching (PTM) method [42]. The threshold in the PTM model is equal to 0.15. The threshold sets an upper limit on the maximum permitted deviation before a local structure is identified as disordered. PTM determines the local lattice structure around an atom and classifies the structures according to the topology of the local atomic environment [42].

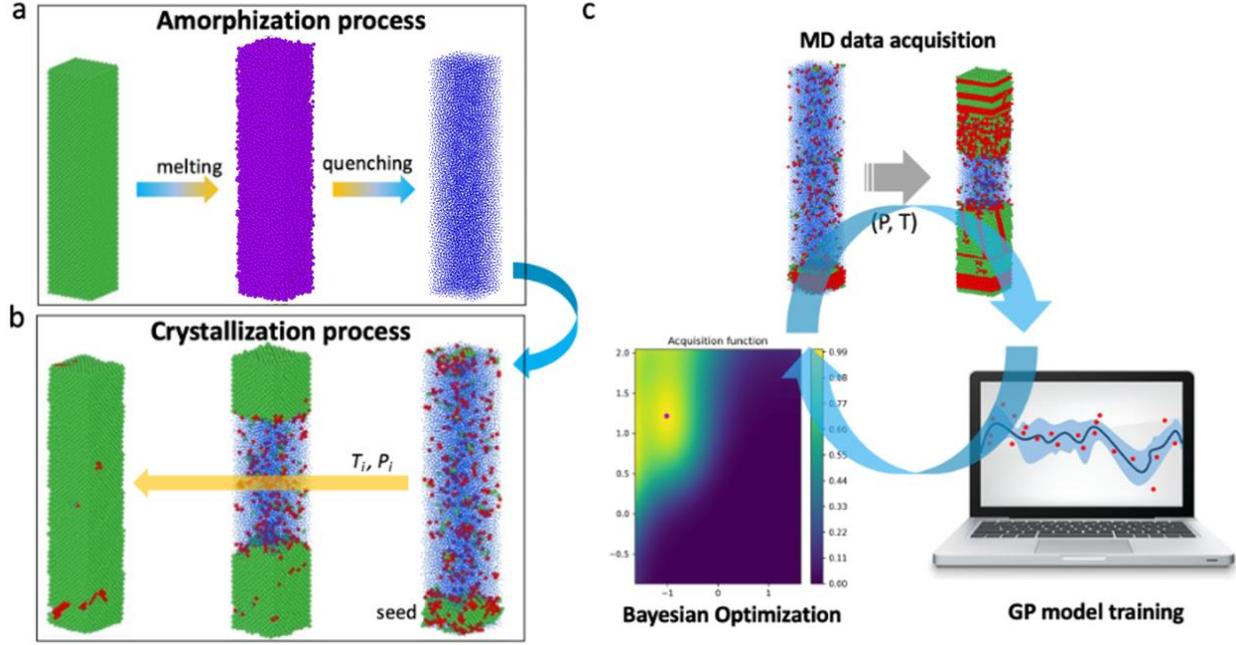

**Figure 1.** (a) Amorphization of fcc-Ni and (b) crystallization of a-Ni. The green, purple, blue and red particles show atoms with fcc, melt, amorphous and hcp phases, respectively. (c) Flowchart of the BO process that iterates between MD data acquisition, GP model training and BO prediction.

*2.2. Overview of the BO process*

**Figure 1c** depicts the BO process that combines MD simulations, GP model training and BO in an iterative loop. For starting the BO, $N$ initial data points as $\mathcal{D}_{1:N} = (\mathbf{x}_1, y_1), \ldots, (\mathbf{x}_N, y_N)$ are produced from MD simulations at different conditions, where $\mathbf{x}_1 = (T_1, P_1)$, $\ldots$, $\mathbf{x}_N = (T_N, P_N)$, and they have been chosen randomly from grid-like points distributed in the $T$-$P$ space. The $y_1, \ldots, y_N$ are the hcp-Ni fractions at steady state in the MD simulations corresponding to $\mathbf{x}_1$, $\ldots$, $\mathbf{x}_N$, respectively. The initial input data are standardized using $\mathbf{x} = \frac{\mathbf{X}_{in} - \mu}{\sigma}$ before feeding them into the multivariate GP model with $\mathbf{X}_{in}$, $\mu$ and $\sigma$ as the input data, mean and standard deviation, respectively. Using the GP model, the posterior mean and standard deviation are calculated and then they are fed into the expected improvement acquisition function, $\alpha(\mathcal{D}_{1:N}, \mu(\mathbf{x}), \sigma(\mathbf{x}), \delta)$. Through the optimization of $\alpha$ with the GP model, the next sampling point is predicted as $\mathbf{x}_{N+1} = argmax_{\mathbf{x}}\, \alpha(\mathbf{x}|\mathcal{D}_{1:N})$. $\mathbf{x}_{N+1}$ is converted to $\mathbf{X}_{N+1}$ by inverting the standardization process, and MD simulation is performed at condition $\mathbf{X}_{N+1}$ to compute the hcp-Ni fraction for this BO-suggested point as $y(\mathbf{X}_{N+1}) = y_{N+1}$. The new data is added to the previous data as $\mathcal{D}_{1:N+1} = \mathcal{D}_{1:N} \cup (\mathbf{x}_{N+1}, y_{N+1})$, which is used to retrain the GP model. This iteration is repeated while no higher hcp-Ni fraction can be found. BO, GP and the acquisition function are explained in details in **section S2** of the SI.

## 3. Result and discussion

**Figure 2a** shows the variation of the hcp-Ni fraction as a function of time during the crystallization of a-Ni with an hcp seed at 800 K and different pressures. The results for two other temperatures at 600 K and 1200 K are shown in **section S3.1** of the SI. In general, for a given temperature, higher hcp-Ni fraction can be reached with higher pressures, but this trend is not monotonic. **Figure 2b** summarizes the steady state hcp-Ni fraction from each condition. The highest hcp-Ni fraction achieved in these simulations is 44% at 600 K and 500,000 bar. We note that for these simulations, the calculation of the hcp phase fraction includes the seed as well.

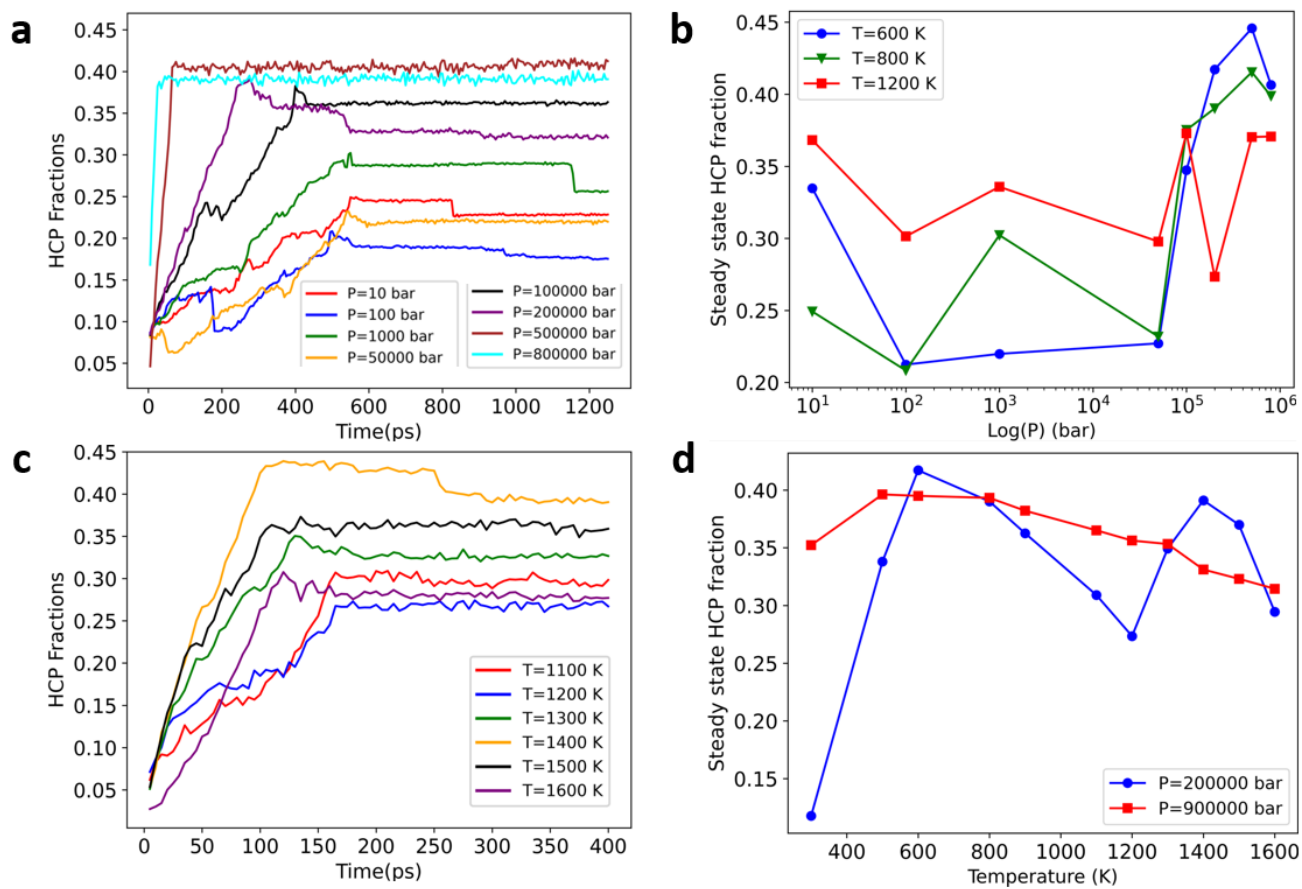

**Figure 2.** **(a)** Fractions of hcp-Ni phase as a function of time for crystallization of a-Ni with a hcp seed at 800 K and different pressures. **(b)** The steady-state hcp-Ni fraction at different temperatures as a function of logarithmic pressure. **(c)** Fractions of hcp-Ni versus time at 200,000 bar and different temperatures. **(d)** The steady-state hcp-Ni fraction at different pressures as a function of temperature.

For two high pressures, we have also performed simulations with different temperatures. **Figure 2c** shows such a series of simulations at 200,000 bar, and those for 900,000 bar is included in **section S3.1** of the SI. The highest steady state hcp-Ni fraction reached in these simulations is 42% at 600 K and 200,000 bar. However, as summarized in **Figure 2d**, there is again no monotonic trend for hcp-Ni fraction as a function of temperature.

Besides employing a hcp crystallite as the crystallization seed, we have also performed the same set of simulations using an fcc seed (**Figure 3** and **section S3.2** in the SI). Similar to the hcp seed case, there are no monotonic trends for the hcp-Ni fraction as a function of temperature nor pressure. The highest hcp-Ni fraction achieved from these simulations is 38% at 900 K and 900,000 bar. For many of the simulations with an fcc seed corresponding to the low pressures, the steady-state hcp-Ni fractions are close to zero (**Figure 3b** and **3d**). Another interesting observation is that for intermediate pressures of 50,000 - 200,000 bar, there is first a gradual increase in the hcp-Ni fraction followed by a sudden decrease to almost zero due to the conversion of the hcp-Ni phase into the fcc-Ni phase. The insets in **Figure 3a** show the representative phases before and after this conversion. The same sharp phase conversion is also observed at other two temperatures simulated (600 and 1,200 K in **sections S3.2 and S4** in the SI). This conversion phenomenon appears only at medium pressures for the crystallization of a-Ni with an fcc seed, indicating these pressures allow the system to form the hcp phase but are not high enough to keep the phase. At higher pressures, the formed hcp phase becomes more stable.

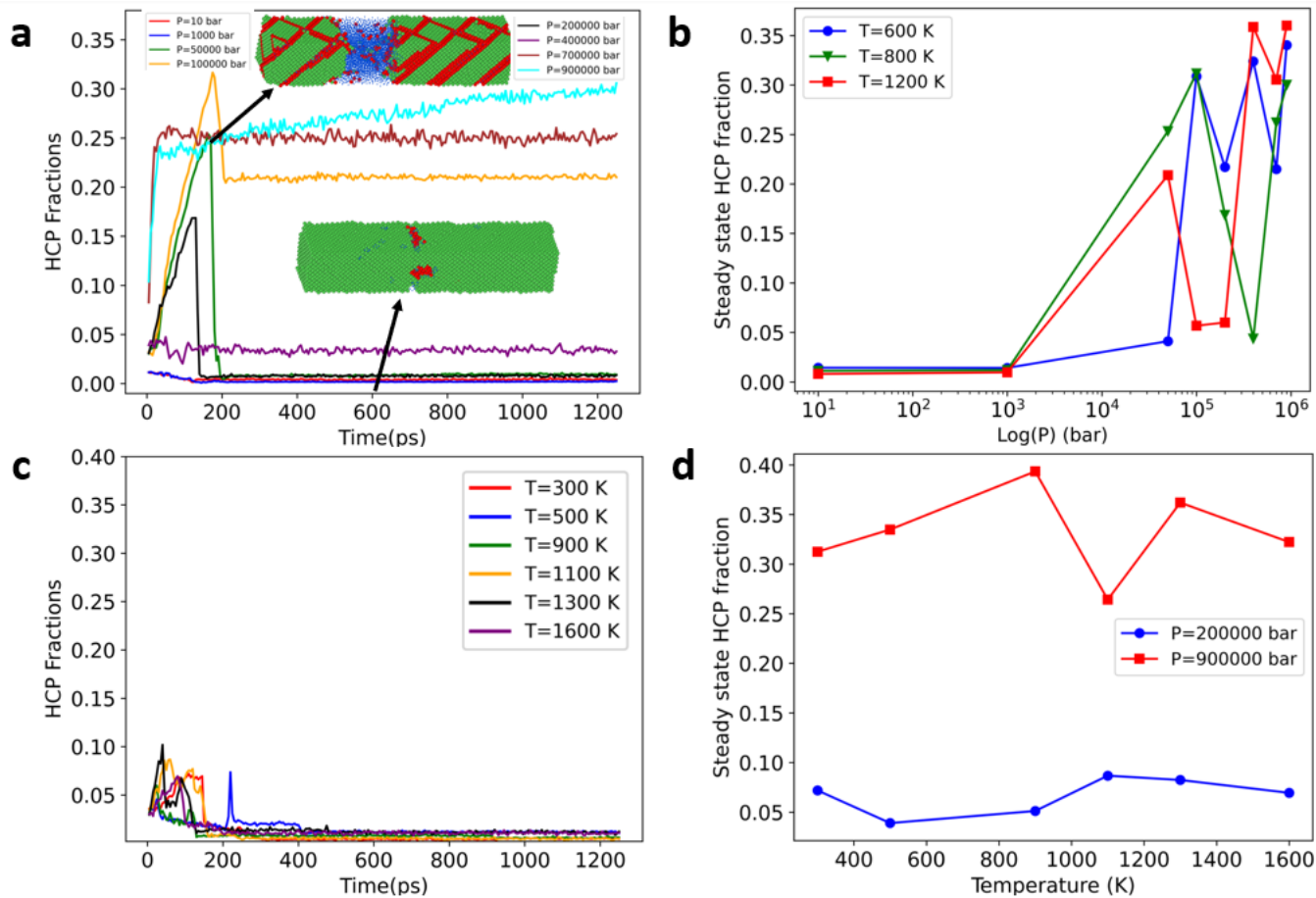

**Figure 3.** **(a)** Fractions of hcp-Ni phase as a function of time for crystallization of a-Ni with an fcc seed at 800 K and different pressures. Insets show representative simulation snapshots before and after the sharp hcp-to-fcc phase transition. **(b)** The steady-state hcp-Ni fraction at different temperatures as a function of logarithmic pressure. **(c)** Fractions of hcp-Ni versus time at 200,000 bar and different temperatures. **(d)** The steady-state hcp-Ni fraction at different pressures as a function of temperature.

The above results indicate that it is challenging to identify the crystallization conditions that may maximize the hcp-Ni fraction due to lack of monotonic trends. Therefore, BO is employed. For BO of the crystallization with an hcp seed, there are 30 data points from the above simulations with different $(T, P)$ pairs. This dataset is used to train the GP model. In **section S2.1.1** in the SI, we show the parity plots corresponding to the Leave-One-Out cross validation [43,44] using our GP model which validates the good predictivity of the model. We bound the variables in the BO as the $[T_{min}, T_{max}] = [300\ K, 2000\ K]$ and $[P_{min}, P_{max}] = [1\ bar, 900000\ bar]$. **Figures 4a-c** are the posterior means, standard deviation and normalized EI acquisition function corresponding to the BO iterations 1, 50 and 85, respectively, for the crystallization with an hcp seed. The noise standard deviation in the GP model is 0.10. For each iteration, one point as the favorable pair of $(T_{new}, P_{new})$ is suggested by BO where the acquisition function is maximized. This suggested condition is simulated using MD to calculate the hcp-Ni fraction, which is then added to the next iteration. With increasing number of iterations, the density of the BO-suggested points in the search space increases where the posterior mean is the highest while the posterior standard deviation is the lowest. In **Figure 4**, as a general view, the leftmost side of the search space includes points that hcp-Ni fraction gets high values. This region corresponds to low temperatures, but the pressures are very high, which makes the crystallization possible.

In **Figure 5a**, the distance between consecutive BO-suggested points in the search space as a function of the iteration number is shown. It is seen that for most of the iterations, the displacements of the suggested points are very small which shows that the suggested points by the acquisition function are so close to each other and the exploitation is dominant to exploration. However, there are some points in the **Figure 5a** that are representations of large distances between consecutive points which is the indication of exploration of the BO, e.g., the distance between pairs of x[22] and x[21] (suggested points at iterations 22 and 21) as (300 K, 374564 bar) and (300 K, 551890 bar), respectively. **Figure 5b** depicts the maximum accumulative values of the hcp-Ni fraction calculated with MD simulations as $y_{best}$ for different BO iterations. Two high values of hcp-Ni happen at iteration numbers 9 and 13 which are equal to 57.44% and 58.25%, respectively. The features corresponding to these two points are the pairs of (300 K, 463186 bar) and (300 K, 459952 bar) which are so close to each other. This result demonstrates that the hcp-Ni fraction for the crystallization of the a-Ni with an hcp seed is very sensitive to pressure values. After 13 iterations, BO could be able to find the maximum value of the hcp-Ni as 58.25%.

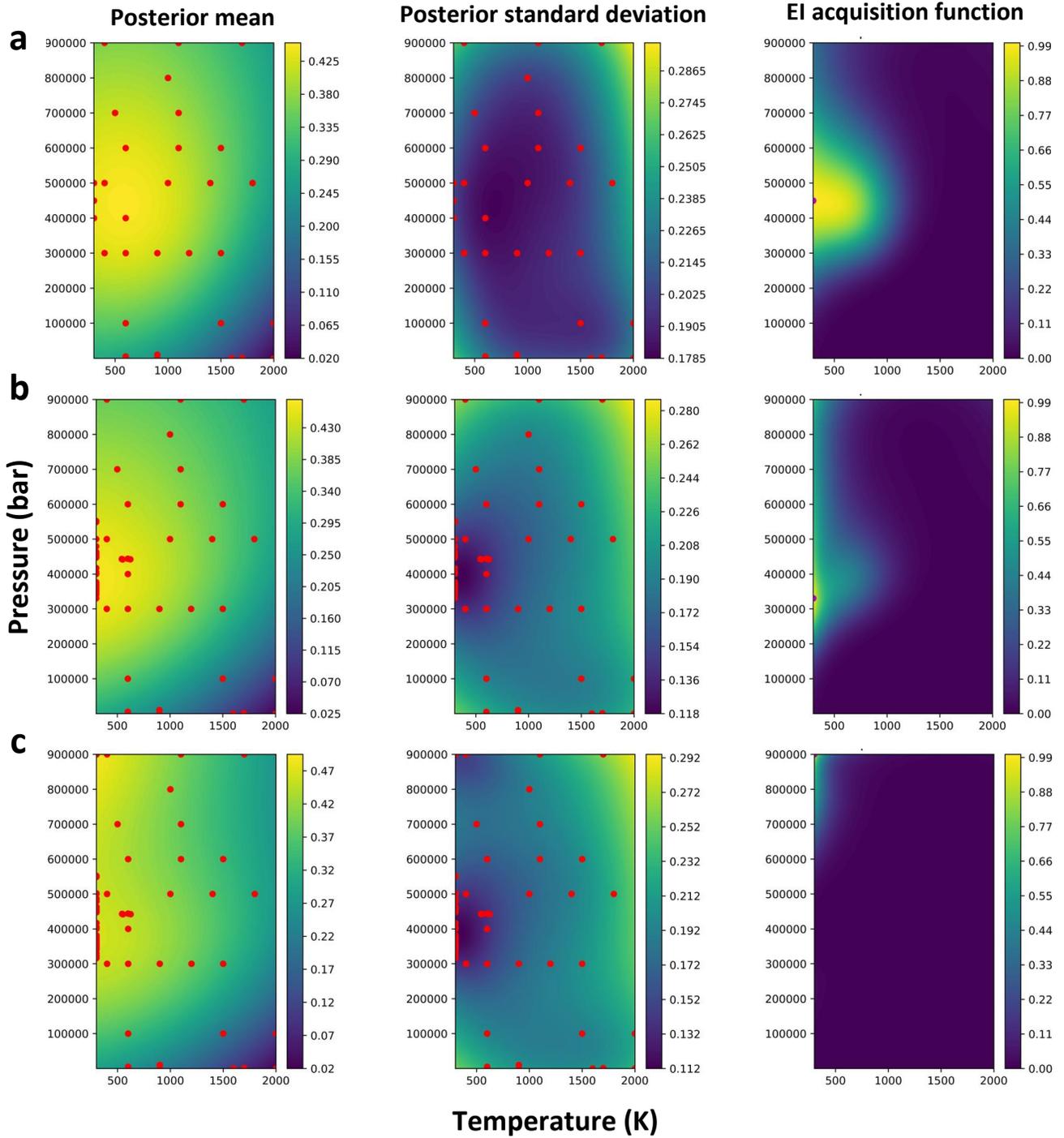

**Figure 4.** Posterior mean, posterior standard deviation and normalized EI acquisition function for the BO of hcp-Ni fraction from crystallization with an hcp seed for iterations **(a)** 1, **(b)** 50, and **(c)** 85.

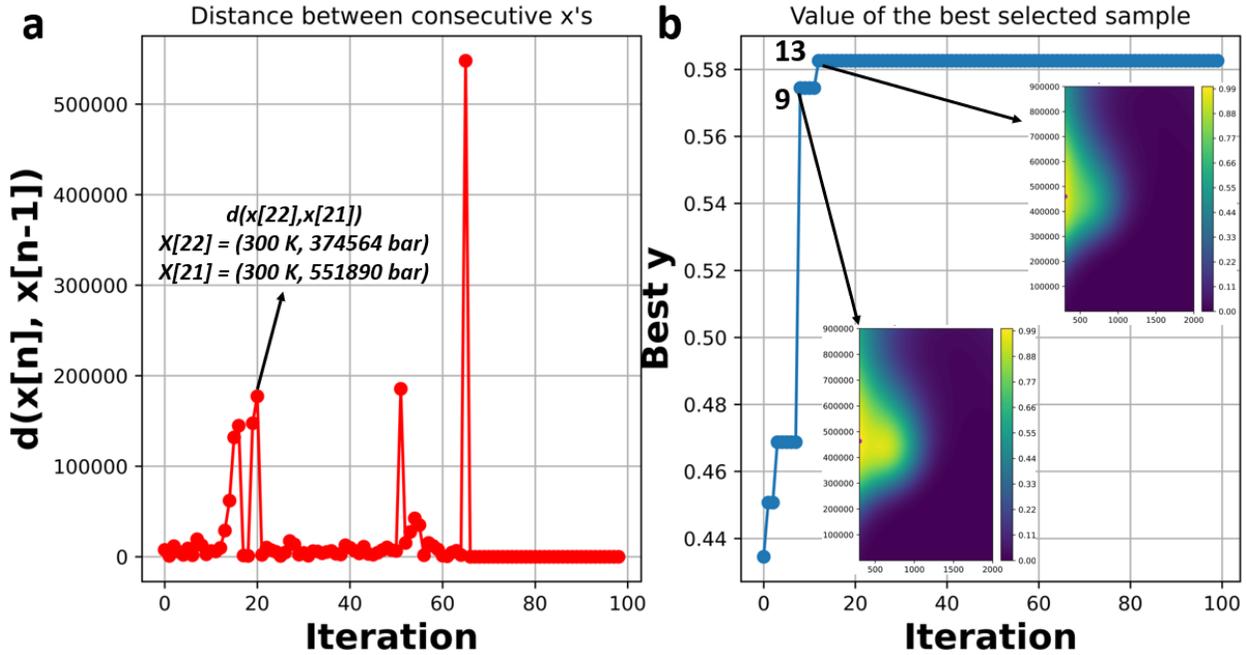

**Figure 5. (a)** The changes of distance between consecutive points suggested by the acquisition function corresponding to the iterations n and n-1 with iteration number **(b)** The changes of best observation calculated using the MD simulation with the iteration number. Two subfigures show the acquisition function distribution for iterations 9 and 13. $\sigma_y$ and $\delta$ are equal to 0.1 and 0.01 for the BO task.

**Figures 6a-c** are the posterior means, standard deviation and normalized EI acquisition function corresponding to the BO iterations 1, 50 and 85, respectively, for the crystallization with an fcc seed. The noise standard deviation in the GP model is 0.03. For BO of the crystallization with an fcc seed, there are 42 initial data points. In **Figure 6**, as a general view, the center of the search space includes most of the points suggested by acquisition function.

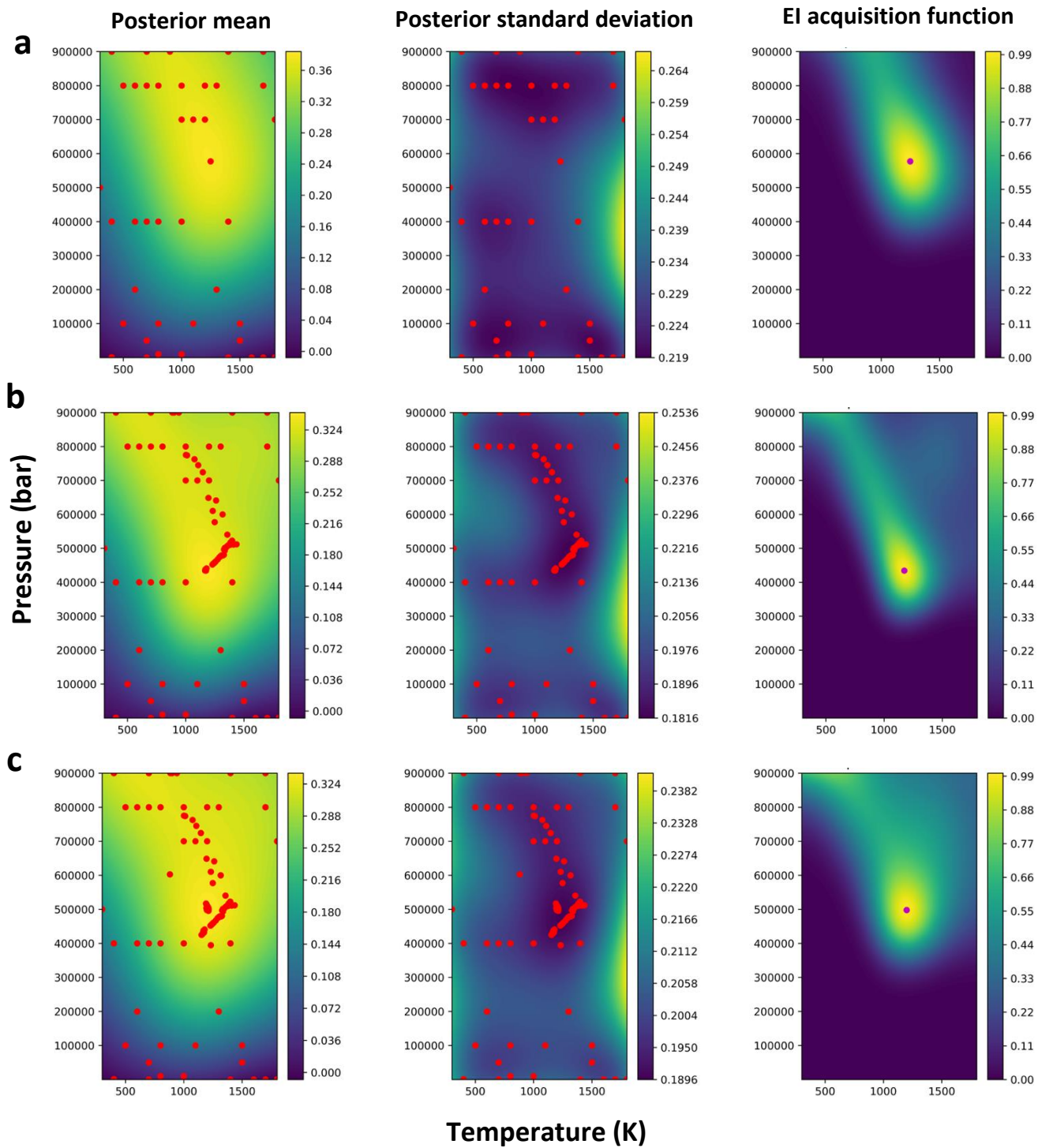

**Figure 6.** Posterior mean, posterior standard deviation and normalized EI acquisition function for the BO of hcp-Ni fraction from crystallization with an fcc seed for iterations **(a)** 1, **(b)** 50, and **(c)** 85.

In **Figure 7a**, the distance between consecutive BO-suggested points in the search space as a function of the iteration number is shown. It is seen that at the beginning, the displacements of the suggested points are large, representing the exploration phase. Later in the BO, the distances between the consecutive points becomes much smaller, indicating the exploitation around the maximum. These can be seen from the acquisition function heat maps from iterations 40 and 77 in **Figure 7b**. Based on these diagrams, it could be seen that the acquisition function distributions are similar to each other and also the suggested red points are close together. In **Figure 7a**, the distance between pairs of x[60] and x[59] (suggested points at iterations 60 and 59) as (1231 K, 394643 bar) and (881 K, 602672 bar), respectively, is large which is the indication of exploration. **Figure 7b** demonstrates two high values of hcp-Ni happening at iteration numbers 40 and 77 which are equal to 43.37% and 42.10%, respectively. The features corresponding to these two points are the pairs of (1229 K, 452358 bar) and (1211 K, 495516 bar). After 40 iterations, BO could be able to find the maximum value of the hcp-Ni as 43.37%.

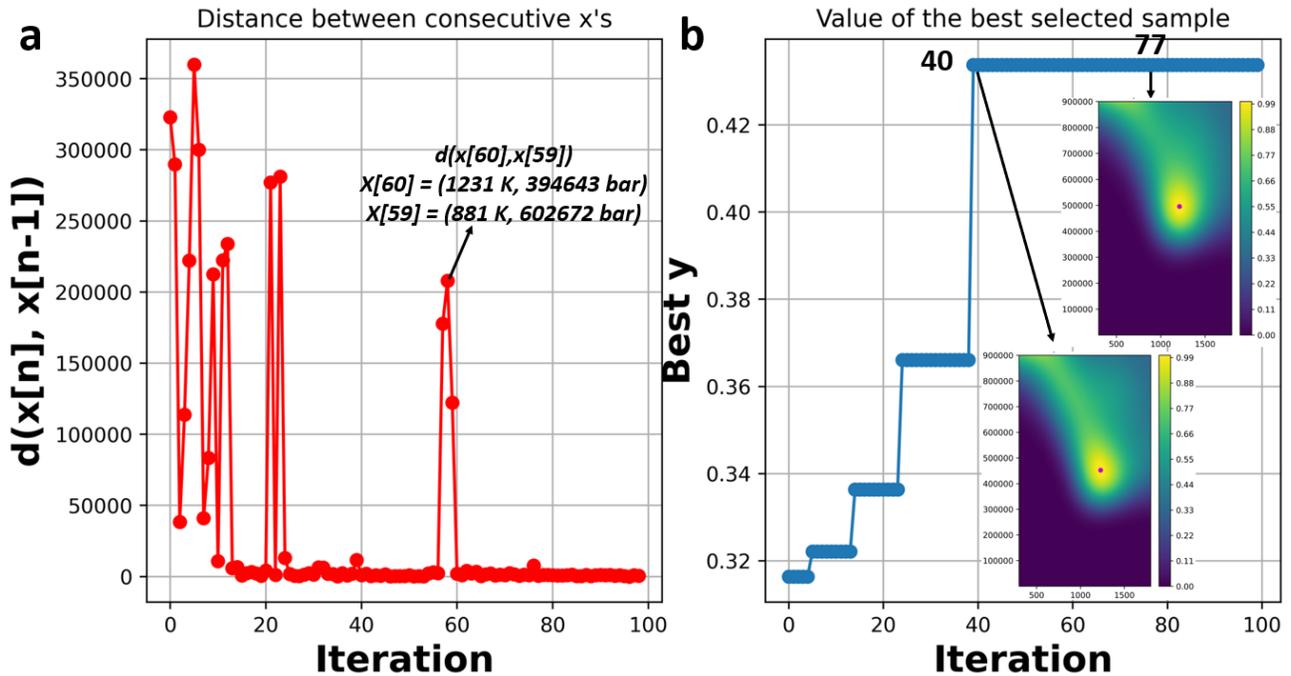

**Figure 7. (a)** The changes of distance between consecutive points suggested by the acquisition function corresponding to the iterations n and n-1 with iteration number **(b)** The changes of best observation calculated using the MD simulation with the iteration number. Two subfigures show the acquisition function distribution for iterations 40 and 77. $\sigma_y$ and $\delta$ are equal to 0.03 and 0.01 for the BO task.

## 4. Conclusion

In this work, the goal is maximizing the fraction of metastable hcp phase during the crystallization of the bulk a-Ni structure at steady state and discover the process conditions leading to that optimum value. Non-equilibrium MD simulations under NPT ensemble are performed to study the crystallization of a-Ni. Using the PTM method, the phases of the atoms as amorphous, fcc and hcp are recognized during the crystallization under different temperature and pressure values and the fractions of hcp phase at steady state are collected as the objective function for the BO algorithm. The surrogate model used in BO algorithm is the GP with matern kernel. Based on the BO of the a-Ni with hcp seed using $\sigma_y$ and $\delta$ equal to 0.1 and 0.01, 58.25% hcp-Ni could be gained after 13 BO iterations where the temperature and pressure are equal to 300 K and 459,952 bar, respectively. Moreover, for the a-Ni with fcc seed using $\sigma_y$ and $\delta$ equal to 0.03 and 0.01, maximum hcp-Ni that BO converges into is 43.37% after 40 iterations of BO algorithm. This study shows the promise of using BO to identify the process conditions that can maximize the rare phases. This method can also be generally applicable to process optimization to achieve target material properties.

## Supplementary Material

See the supplementary material for the complete explanation of the *Evolution of process conditions*, *Bayesian Optimization*, *Variation of hcp-Ni fraction with time* and *Crystallization of a-Ni with FCC seed at medium pressure values*.

## Code availability

The Python code required to reproduce these findings are available to download from https://github.com/sinaDFT/BO_BulkNi_Crystallization upon publication.

## Acknowledgements

The work was performed with financial support by U.S. Army Research Office Grant # W911NF2110045 under the Materials Synthesis & Processing Program, with Dr. Michael P. Bakas as the program manager. The computations are supported by the University of Notre Dame, Center for Research Computing (CRC).